\documentclass[doublecol,amssymb,aps]{epl2} 

\usepackage{graphicx}
\usepackage{dcolumn}
\usepackage{amsmath}
\usepackage{bm}
\usepackage{color}

\usepackage{hyperref}
\usepackage{txfonts}
\usepackage{pxfonts}
\usepackage{graphicx,bm,units,yfonts}

\newcommand{\CM}{{\mathbb C}}

\newcommand{\RM}{{\mathbb R}}

\newcommand{\TM}{{\mathbb T}}
\newcommand{\ZM}{{\mathbb Z}}

\newcommand{\EM}{{\mathbb E}}

\newcommand{\Ss}{{\cal S}}
\newcommand{\Oo}{{\cal O}}
\newcommand{\Tt}{{\cal T}}

\newcommand{\one}{{\bf 1}}

\newcommand{\Tr}{\mbox{\rm Tr}}
\newcommand{\ev}{{\mbox{\tiny\rm ev}}}
\newcommand{\od}{{\mbox{\tiny\rm od}}}

\newcommand{\Ch}{{\rm Ch}} 
 
\newcommand{\Ker}{{\rm Ker}}

\newcommand{\Sig}{{\rm Sig}}

\newcommand{\Red}{}

\title{Invariants of disordered semimetals via the spectral localizer}
\shorttitle{Invariants of disordered semimetals via the spectral localizer} 

\author{Hermann Schulz-Baldes  \and Tom Stoiber}
\shortauthor{Schulz-Baldes and Stoiber}

\institute{ 
Department Mathematik, FAU Erlangen-N\"urnberg, Cauerstr. 11, D-91058 Erlangen, Germany                   
}

\abstract{
The spectral localizer consists of placing the Hamiltonian in a Dirac trap. For topological insulators its spectral asymmetry is equal to the topological invariants, providing a highly efficient tool for numerical computation. Here this technique is extended to disordered semimetals and allows to access the number of Dirac or Weyl points as well as weak invariants. These latter invariants imply the existence of surface states.
}

\begin{document}

\maketitle

\section{Introduction}

The defining feature of periodic semimetals is the existence of Dirac or Weyl points at the Fermi level \cite{NGM,WTVS}. In the vicinity of such a point the dispersion relation is approximately linear and can be described by a Dirac or Weyl Hamiltonian. This leads to a pseudogap in the density of states at the Fermi level. In dimension $d=2$, Dirac points are generic and stable for periodic Hamiltonians having a sublattice (chiral) symmetry, the prototypical example being graphene \cite{NGP}. On the other hand, for dimension $d=3$ Weyl points are well-known to be generic for periodic Hamiltonians by the von Neumann-Wigner theorem \cite{NW}, but fixing them to the Fermi level requires further symmetries \cite{AMV}. Non-generic situations such as nodal-line semimetals are not further considered in this work. Each Dirac and Weyl point is a monopole of Berry curvature and hence has an associated a topological charge given by integrating the Berry curvature over a $d$-sphere around it. This topological nature is tightly linked to the fact that Dirac and Weyl points appear at transitions between different phases of topological insulators \cite{Mur,PSbook,Van,MM}.

A hallmark of semimetals is the existence of surface states. For graphene, this has been known for more than two decades  \cite{NDD,DUM,FLW}, and for Weyl semimetals the surface states were termed Fermi arcs \cite{WTVS,Bal,XWW,MT}. Both have been accessible in experiments \cite{NGP,LWF}. Another physical effect in $3d$ Weyl semimetals is the anomalous Hall effect \cite{YLR,BB,Tak}. Furthermore, the number of Weyl points is linked to an electro-magnetic response \cite{NN,Bur,AMV} (see \cite{ZMS} about doubts on that though) as well as a quantized circular photogalvanic effect \cite{JGMM}.  The stability of semimetals and the associated physical effects as described above w.r.t. disordered perturbations remains a disputed issue \Red{\cite{Tak,AB,NandkishoreEtAl2014,PHD,SlagerEtAl2017,BDA,PW21}}.

The main objective of this work is to provide new tools for the numerical analysis of Dirac and Weyl points as well as other topological quantities in possibly dirty semimetals. This is achieved by adapting the so-called spectral localizer \cite{LS1,LS2} to the study of semimetals. The number of Dirac or Weyl points is shown to be equal to the \Red{dimension of the approximate kernel} of the spectral localizer in a semiclassical limit. In particular, this allows to extend the notion of Dirac and Weyl points to disordered semimetals. Furthermore, another version of the spectral localizer allows to access weak invariants which guarantee the existence of surface states.

\section{Spectral localizer for topological insulators}

As it is needed to explain the new applications to semimetals, let us begin by recalling the basic facts about the spectral localizer for gapped topological insulators. In two recent works, Terry Loring and one of the present authors showed that the strong topological invariants of non-interacting fermionic topological insulators can be computed using the so-called spectral localizer \cite{LS1,LS2}. \Red{Suitable modifications allow to compute weak invariants \cite{FPL,LSM,SSt0},} spin Chern numbers and alike \cite{DoS} as well as $\ZM_2$-invariants \cite{DoS2}. In these previous works there are two basic versions of the spectral localizer, pending on whether one wants to compute even or odd invariants (respectively Chern numbers or winding numbers which are also called odd Chern numbers). In both versions it is a self-adjoint operator constructed from a tight-binding Hamiltonian placed in a Dirac trap. If the Hamiltonian is insulating in the bulk, the spectral localizer with open boundary conditions defines a finite-dimensional matrix with a stable spectral gap even in the presence of edge modes and then its half-signature measuring its spectral asymmetry is equal to the topological invariant. Compared to other approaches such as twisted boundary conditions \cite{EM}, the Bott-Index \cite{LH} or real-space methods \cite{Pro}, this is a very efficient and flexible numerical method for the computation of topological invariants. It also has the advantage that it can work directly with {\it e.g.} quasi-periodic or amorphous systems where boundary conditions are difficult or impossible to apply. \Red{Another particularly fast real space method is the kernel polynomial method, but it seems to be limited to integer-valued invariants \cite{VarEtAl}.}

For a more technical description, let us first focus on the case of even dimension $d$ \cite{LS2,LSS}. Let $H$ be a generic short-range tight-binding Hamiltonian with $L$ matrix degrees of freedom, hence acting on the Hilbert space $\ell^2(\ZM^d,\CM^L)$. Its hopping terms and potentials are supposed to form a covariant family ({\it e.g.} random or quasiperiodic, see \cite{PSbook} for a technical definition) and be equipped with an average over the configurations which is denoted by $\EM$.  The Fermi level is supposed to lie in a gap of size $g>0$ and is, after a suitable shift of energy, equal to $0$. Then the even spectral localizer for coupling parameter $\kappa>0$, setting the non-commutative scale of space quanta, is
\begin{equation}
\label{eq-EvLocDef}
L^\ev_\kappa
\;=\;
\begin{pmatrix}
-H & \kappa\,D_0^* \\ \kappa\,D_0 & H
\end{pmatrix}
\;,
\qquad
D\;=\;
\begin{pmatrix}
0 & D_0^* \\ D_0 & 0
\end{pmatrix}
\;,
\end{equation}
where the dual Dirac operator $D$ is built from the components $X_1,\ldots, X_d$ of the position operator
\begin{equation}
\label{eq-Dirac}
D\;=\;
\sum_{j=1}^d \gamma_j X_j
\;,
\end{equation}
with $\gamma_1,\ldots,\gamma_{d}$ being an irreducible selfadjoint representation of the Clifford algebra $\CM_d$ on $\CM^{d'}$ with $d'=2^{\lfloor\frac{d+1}{2}\rfloor}$ such that $\gamma_d$ is the Pauli matrix $\sigma_2$, and, for even $d$, $\gamma_{d+1}=\sigma_3$ induces $\gamma_{d+1}D\gamma_{d+1}=-D$. Hence $L^\ev_\kappa$ is a selfadjoint operator on $\ell^2(\ZM^d,\CM^{Ld'})$ with compact resolvent. The finite volume restriction (with Dirichlet boundary conditions) of the spectral localizer to the range of $\chi(D^*D\leq \rho^2)$ is denoted by $L^\ev_{\kappa,\rho}$. The main result of \cite{LS2} states that, provided that
\begin{equation}
\label{eq-BoundHyp}
\kappa\;<\;\frac{12\, g^3}{\|H\|\,\|[D_0,H]\|}
\qquad\mbox{and}
\qquad
\rho\;>\;\frac{2\,g}{\kappa}
\;,
\end{equation}
the finite-volume spectral localizer $L^\ev_{\kappa,\rho}$ has a gap of size at least $g/2$ and its signature (number of positive minus number of negative eigenvalues) determines the $d$th Chern number of the Fermi projection $P=\chi(H<0)$:
\begin{equation}
\label{eq-ChSig}
\Ch_d(P)
\;=\;
-\,\frac{1}{2}\,\Sig(L^\ev_{\kappa,\rho})
\;.
\end{equation}
Here the $d$-dimensional Chern number is by definition
\begin{align*}
{\mathrm{Ch}}_{d}(P) 
\;=\; 
\frac{(2\imath \pi)^{\frac{d}{2}}}{ \frac{d}{2}!} 
\,\sum_{\nu\in S_d}(-1)^\nu
\,\EM\,\mathrm{Tr}\, 
\langle 0| P \prod_{j=1}^{d} \imath [X_{\nu(j)},P]|0\rangle  \; ,
\end{align*}
where $S_d$ is the symmetric group of permutations and $(-1)^\nu$ denotes the signature of a perturbation $\nu$. This is the so-called strong topological invariant and an index theorem assures that it is integer-valued \cite{PSbook}. The first condition in \eqref{eq-BoundHyp} can always be satisfied for a short range Hamiltonian (so that $\|[D_0,H]\|<\infty$) by choosing $\kappa$ sufficiently small, which means that the resolution of non-commutative space is sufficiently fine. Note also that the bound is invariant under scaling $H\mapsto \lambda H$ and $L^\ev_\kappa \mapsto \lambda L^\ev_{{\kappa}/{\lambda}}$. The second condition in \eqref{eq-BoundHyp} can then be satisfied by choosing $\rho$ sufficiently large. For a Hamiltonian with gap of order $1$, $\kappa\approx 0.1$ and $\rho\geq 10$ is typically already sufficient. 

The numerical advantage of computing $\Ch_d(P)$ by \eqref{eq-ChSig} is rooted in the fact that the definition \eqref{eq-EvLocDef} of the spectral localizer does not involve any functional calculus of the Hamiltonian which is necessary if the definition of ${\mathrm{Ch}}_{d}(P)$ is implemented directly \cite{Pro}. It is remarkable, but not proven, that the relation \eqref{eq-ChSig} extends to the mobility gap regime \cite{LSS}.

Next let us briefly describe the odd spectral localizer used for the computation of (higher) winding numbers associated to chiral Hamiltonians in odd dimension $d$ \cite{LS1}. The gapped Hamiltonian is supposed to have a chiral symmetry $H=-\sigma_3 H\sigma_3$ w.r.t. the third Pauli matrix. With $D$ as in \eqref{eq-Dirac}, the odd spectral localizer is
\begin{equation}
\label{eq-OdLocDef}
L^\od_\kappa
\;=\;
\begin{pmatrix}
 \kappa\,D & A^*\\ A & -\kappa\,D
\end{pmatrix}
\;,
\qquad
H\;=\;
\begin{pmatrix}
0 & A^* \\ A & 0
\end{pmatrix}
\;.
\end{equation}
Again the bounds \eqref{eq-BoundHyp}, with $D_0$ replaced by $D$, insure the invertibility of $L^\od_\kappa$ and then
\begin{equation}
\label{eq-OdChSig}
\Ch_d(A)
\;=\;
\frac{1}{2}\,\Sig(L^\od_{\kappa,\rho})
\;,
\end{equation}
where the integer-valued (higher) winding numbers (odd Chern numbers) are defined by \cite{PSbook}
$$
\Ch_{d}(A) 
\;=\; 
\frac{\imath(\imath \pi)^\frac{d-1}{2}}{ d!!} \sum_{\nu\in \Ss_d} (-1)^\nu \,\EM\,\Tr\,\langle 0|\prod_{j=1}^{d}   A^{-1} \imath [A,X_{\nu(j)}] |0\rangle
\;.
$$
Often this is written using the Fermi unitary $U=A|A|^{-1}$ for which $\Ch_{d}(U)=\Ch_{d}(A)$.

\section{Counting Weyl and Dirac points}

Let us now turn to semimetals (without interaction) which are also supposed to be described by a tight-binding Hamiltonian $H$ on $\ell^2(\ZM^d,\CM^L)$. As already stated in the introduction, periodic semimetals have by definition Dirac or Weyl points at the Fermi level (still set to $0$). Locally in $k$-space, they can be approximated by a Hamiltonian on $L^2(\RM^d,\CM^{d'})$ given by
\begin{equation}
\label{eq-DiracWeyl}
H^{D/W}\;=\;\sum_{j=1}^d a_j\,\Gamma_j\,\tfrac{1}{\imath}\,\partial_{k_j}
\;,
\end{equation}
where $a_j\not=0$ are the slopes of the energy bands at the Dirac or Weyl points, $\partial_{k_j}$ are the partial derivatives and $\Gamma_1,\ldots,\Gamma_d$ is another irreducible selfadjoint Clifford algebra representation on $\CM^{d'}$. For $d$ even, one has a Dirac Hamiltonian satisfying $H^D=-\Gamma_{d+1}H^D\Gamma_{d+1}$, while for odd $d$ one speaks of a Weyl Hamiltonian $H^W$. Degenerate band-touching points in odd dimension that are composed of an equal number of left- and right-handed Weyl Hamiltonians are commonly also referred to as Dirac points (which slightly deviates from the terminology used here). Both Dirac and Weyl points have a pseudogap in the density of states at $0$. If $H$ has no other band at $0$ it inherits this pseudogap. The topological charge of each Dirac or Weyl point is obtained by integrating the monopole of Berry curvature over a ball around the Dirac or Weyl point and is determined by the signs of the coefficients $a_j$. Isolated Weyl points are stable in dimension $d=3$ and can only be eliminated by fusing them with another Weyl point of opposite chirality. In contrast, Dirac points are generically unstable unless protected by additional symmetry, as they can be gapped by a term proportional to $\Gamma_{d+1}$. The total charge summed over the Brillouin zone always vanishes in odd dimension due to the Nielsen-Ninomiya theorem \cite{NN}. The same often also happens in even dimension since stable Dirac points tend to occur in symmetry-related pairs (a counterexample to this rule are unpaired Dirac nodes as they often appear at gap-closings, e.g. at the phase boundaries of the Haldane model). Instead of the charge one might therefore want to consider
$$
I
\;=\;
\mbox{\rm total number of Dirac/Weyl points}\;,
$$
where each Dirac/Weyl point at the Fermi level is counted with its multiplicity. More precisely, one counts the multiplicity of the model Hamiltonians \eqref{eq-DiracWeyl} in an expansion around the band-touching points. Hence two Weyl fermions with opposite chirality forming a minimally degenerate so-called "Dirac point" in three dimensions would give $I=2$.

\begin{figure}[t]
\includegraphics[width=1.0\linewidth]{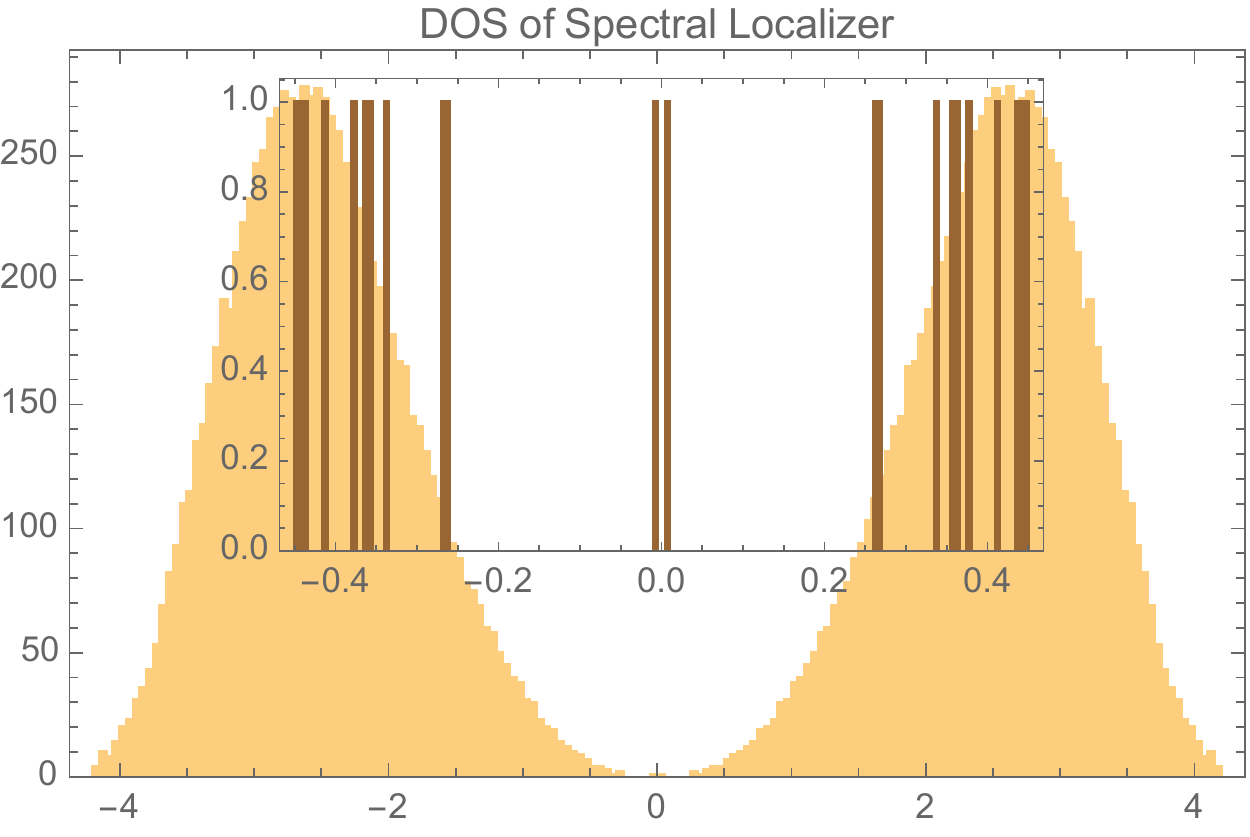}
\caption{Histogram of the eigenvalues of one random realization of $L_{\kappa,\rho}$ for $\kappa = 0.07$ and $\rho=34$, and a stacked SSH Hamiltonian with parameters $\lambda = 0.6$, $\delta = 0.3$ and $\mu = 1.3$. The \Red{dimension of the approximate kernel} is $2$ so that there are $2$ Dirac points. Note that the distance to the next eigenvalues is of order $\kappa^{\frac{1}{2}}\approx 0.26$.}
\label{fig-StackedSSH}
\end{figure}

The first main result of the paper states that $I$ can be computed as the \Red{dimension of the approximate kernel} of a spectral localizer, namely via
\begin{equation}
\label{eq-DWcount}
I\;=\;
\Tr\big(\chi(|L^\od_\kappa|<\epsilon)\big)
\;,
\qquad
L_\kappa
\;=\;
\begin{pmatrix}
 \kappa\,D & H \\ H & -\kappa\,D
\end{pmatrix}
\;,
\end{equation}
for $\epsilon > 0$ fixing a small window around zero whose size will be described further below. Let us stress that, other than in \eqref{eq-EvLocDef} and \eqref{eq-OdLocDef}, the spectral localizer of \eqref{eq-DWcount} satisfies $\sigma_2 L_\kappa\sigma_2=-L_\kappa$ and thus has no spectral asymmetry. It also has no gap (note that \eqref{eq-BoundHyp} does not hold). As will be argued below, the cluster of eigenvalues forming the approximate kernel is nevertheless well-separated from the remainder of the spectrum (see Fig.~\ref{fig-StackedSSH}). To give an intuitive argument why \eqref{eq-DWcount} holds in odd dimensions, one should think of $L_\kappa$ as the spectral localizer for the auxiliary chiral Hamiltonian $\sigma_1 \otimes H$ which has an evenly degenerate Dirac point for each Weyl point of $H$ at the Fermi level. Hence the auxiliary Hamiltonian can be gapped by arbitrarily small chiral perturbations, such as the chiral mass term $m \sigma_2\otimes \one$. Since the resolution of a Dirac-point locally causes the odd Chern number to jump by $\pm 1$ \cite{GJK,PSbook}, one can carefully construct perturbations which result in topological insulators with small gap and odd Chern number ranging between two integers $c_-$ and $c_+$ which differ by precisely $c_+-c_- = I$. Since $\frac{1}{2}\Sig(L_{\kappa,\rho})$ tracks those changes, there should be at least $I$ small eigenvalues whose sign can be flipped by a small perturbation. 

A rigorous argument that \eqref{eq-DWcount} holds for a periodic semimetal in arbitrary dimension while also giving precise asymptotics for the gap to the remainder of the spectrum is obtained by semiclassical analysis in $\kappa$. This shall be sketched in the following. Going to Fourier space, the square $(L_\kappa)^2$ of the spectral localizer becomes
$$
-\,\kappa^2 \sum_{j=1}^{d}\partial^2_{k_j}
\;+\;
\begin{pmatrix}
\widehat{H}_k^2 & -\,\imath\,\kappa\sum_{j=1}^d\gamma_j (\partial_{k_j}\widehat{H}_k) \\
\imath\,\kappa\sum_{j=1}^d\gamma_j (\partial_{k_j}\widehat{H}_k) & \widehat{H}_k^2
\end{pmatrix}
\;,
$$
where $\widehat{H}_k\in\CM^{L\times L}$ is the Bloch Hamiltonian. This is hence a matrix-valued Schr\"odinger operator on the Brillouin torus $\TM^d$ with semiclassical parameter $\hbar=\kappa$. It turns out to have potential wells precisely at the Dirac or Weyl points. One can then show that for each well  there is precisely one zero-mode and that the first excited state  is at $\kappa$. More precisely, the spectral localizer associated to \eqref{eq-DiracWeyl}
$$
L^{D/W}_\kappa
\;=\;
\begin{pmatrix}
 \kappa\,D & H^{D/W} \\ H^{D/W} & -\kappa\,D
\end{pmatrix}
\;,
$$
viewed as selfadjoint operator on $L^2(\RM^d,\CM^{(d')^2})$, has a simple kernel and the first excited state of of $(L^{D/W}_\kappa)^2$ is of order $\kappa$. Of course, the wells interact by tunnel effect, but as usual this spreads the kernel of $(L_\kappa)^2$ only to order $e^{-1/\kappa}$. Summing all contributions shows that \eqref{eq-DWcount} holds for $\epsilon\ll\kappa^{\frac{1}{2}}$. For a disordered system, one can consider \eqref{eq-DWcount} as a noncommutative definition of the number $I$ of Dirac or Weyl points. If $\lambda\geq 0$ denotes the disorder strength, then first order perturbation theory (using the fact that the zero modes of the spectral localizer are approximately Gaussians of variance $\kappa^{-1}$) and the central limit theorem show that \eqref{eq-DWcount} holds for $\epsilon\ll \kappa^{\frac{d}{4}}\lambda$. A more technical proof of all these facts will be provided elsewhere.

\section{Numerical illustration of Dirac and Weyl point count}

A great advantage of \eqref{eq-DWcount} is that it can be implemented numerically in an efficient manner. Let us first consider a chiral Hamiltonian in $d=2$ given by a stacked SSH model (SSH for Su-Schrieffer-Heeger, {\it e.g.} \cite{PSbook}):
$$
H
\,=\,
\begin{pmatrix}
0 & \!\!\!\!\!\! S_1-\mu-\delta(S_2+S_2^*)+\lambda v_n
\\
S^*_1-\mu-\delta(S_2+S_2^*)+\lambda v_n \!\!\!\!\!\! & 0
\end{pmatrix}
\,.
$$
Here $S_1$ and $S_2$ are the shifts on the lattice (namely the hopping terms), $\mu,\delta$ real parameters and $v_n$ independent random variables uniformly distributed in $[-\frac{1}{2},\frac{1}{2}]$. The idea behind this construction is that $\mu+\delta(S_2+S_2^*)$ acts as a varying mass term closing the gap of the one-dimensional SSH Hamiltonians, thus leading to Dirac points.  For $\lambda=0$, the above periodic model has Dirac points if $2\delta\cos(k_2)+\mu=\pm 1$. For $4\delta<2$ and $\mu\in(-1-2\delta,1+2\delta)$, there are $2$ Dirac points, while for $4\delta>2$ and $\mu$ sufficiently small, there are $4$ Dirac points. A typical numerical result illustrating \eqref{eq-DWcount} is given in Fig.~\ref{fig-StackedSSH}. The periodic stacked SSH Hamiltonian with two Dirac points can be deformed into the standard graphene Hamiltonian with the set of $2d$ chiral semimetals, hence specifying the same class of topological semimetal.

Also Weyl point counts in $d=3$ are readily accessible on a laptop. The above procedure of stacking lower dimensional topological insulators to build semimetals can be applied for all $d$. Here let us use two-dimensional $(p+ip)$-model $H_{p+ip}$ in the $1-2$ plane to build a $3d$ Weyl semimetal described by
$$
H_{p+ip}\otimes \one
\:+\;
\delta\,\one\otimes (S_3+S_3^*)
\;+\;
\lambda H_{\mbox{\rm\tiny dis}}
\,,
$$
where $H_{\mbox{\rm\tiny dis}}$ is a generic disordered potential. Implementing this Hamiltonian on a cube of size $15^3$ leads to plots that are very similar to Fig.~\ref{fig-StackedSSH} and from which the number of Weyl points can readily be read off. The Weyl point count even works when the Weyl points are energetically shifted so that there is no pseudo-gap.

\begin{figure}[t]
\includegraphics[width=1.0\linewidth]{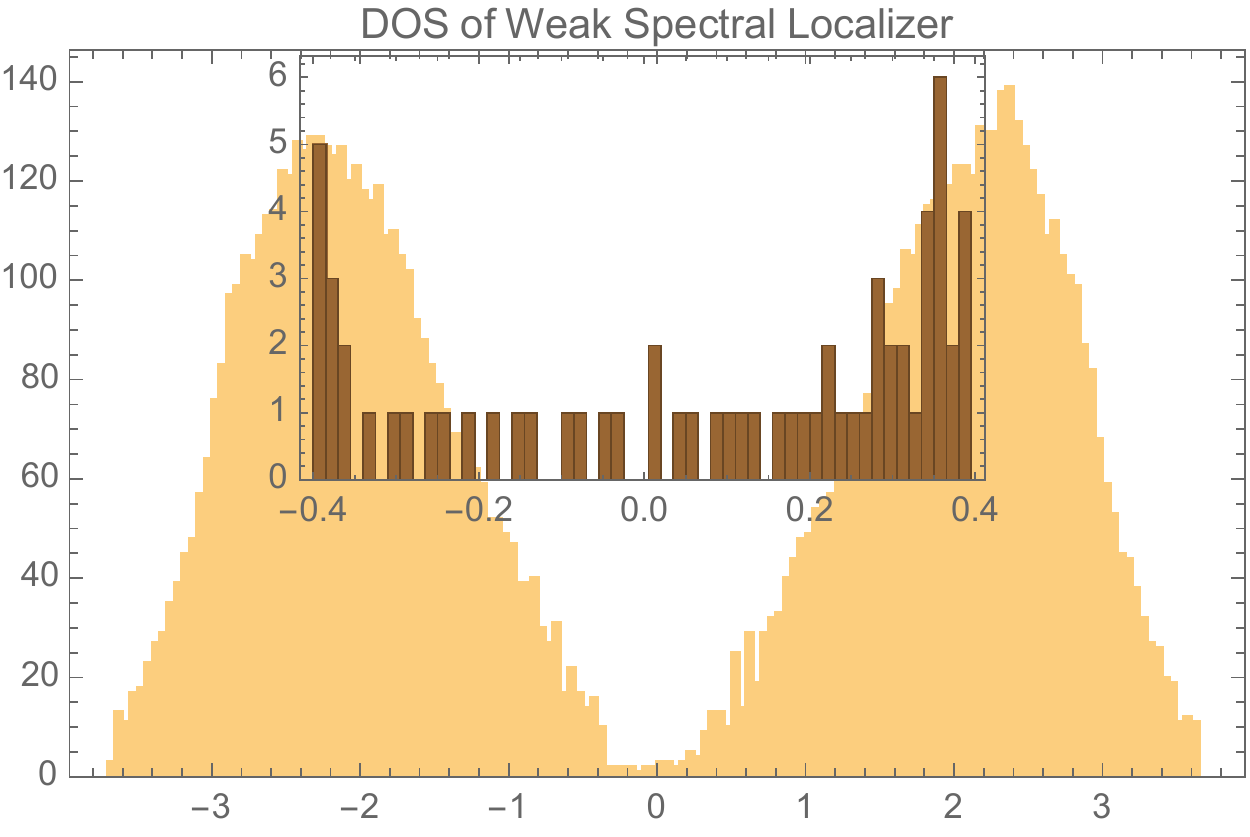}
\caption{Histogram of the eigenvalues of $L^{w,1}_{\kappa,\rho}$. All parameters are as in Fig.~\ref{fig-StackedSSH}. The computed half-signature is $24$ so that $\Ch_{\{1\}}(A)\approx \frac{24}{69}$ because the side length of the sample is $2\rho+1=69$.}
\label{fig-StackedSSHWeak}
\end{figure}

\section{Weak invariants in semimetals}

In absence of additional symmetries the existence of Dirac and Weyl points generically implies that the semimetal has non-vanishing weak (bulk) invariants, which in turn imply the existence of Fermi arcs of surface states. It is hence of great interest to compute these weak invariants and this can again be achieved by a spectral localizer. Let us focus here on chiral Hamiltonians in $d=2$ with off-diagonal entry $A$ as in \eqref{eq-OdLocDef}. Then there are two weak winding numbers given by
$$
\Ch_{\{j\}}(A) 
\;=\; 
\imath \,\EM\,\Tr\,\langle 0|A^{-1} \imath [A,X_{j}] |0\rangle
\;,
\qquad
j=1,2
\;.
$$
Let us stress that $A$ is {\it not} invertible, but the inverse as unbounded operator can have sufficiently regularity for this formula to have a well-defined mathematical meaning. This is the case \cite{SSt} if  the Fermi level either lies in a pseudogap or at least in a region of Anderson localization (in technical terms: the Aizenman-Molcanov bounds \cite{AM} hold). Whether these conditions actually hold for a generic disorder is a heavily disputed question \cite{AB,PHD,Tak,BDA}. For the periodic stacked SSH model described above ($\lambda=0$), one finds that for $\mu>0$, $0<2\delta<1$ and $\mu+2\delta>1$
$$
\Ch_{\{1\}}(A) 
\;=\;
\int^{\pi}_{-\pi}\frac{dk_2}{2\pi}\,\chi\big(-2\delta\cos(k_2)-\mu\geq -1\big)
$$
and $\Ch_{\{2\}}(A) =0$. This shows that the term {\it invariant} is to be taken with a grain of salt, but at least $\Ch_{\{j\}}(A) $ are known to vary continuously in the parameters of the Hamiltonian \cite{SSt}.

To compute the weak winding numbers, one considers the weak spectral localizers as in \cite{SSt0}
$$
L^{w,j}_{\kappa,\rho}
\;=\;
\begin{pmatrix}
\kappa\, X_j & A^*_{\rho,j} \\ A_{\rho,j} & -\kappa\,X_j
\end{pmatrix}
\;,
\qquad
H_{\rho,j} 
\;=\;
\begin{pmatrix}
0 & A^*_{\rho,j} \\ A_{\rho,j} & 0
\end{pmatrix}
$$ 
where $H_{\rho,j}$ is the Hamiltonian on finite volume $[-\rho,\rho]^2$ with Dirichlet boundary condition in direction $j$ and periodic boundary condition in the other. An extension of the results of \cite{SSt0} shows hat
\begin{equation}
\label{eq-WeakComp}
\Ch_{\{j\}}(A) 
\;=\;
\frac{1}{2\rho+1}\,\EM\,\frac{1}{2}\,\Sig(L^{w,j}_{\kappa,\rho})
\;+\;
\Oo(\rho^{-1},\kappa)
\;.
\end{equation}
%
Again \eqref{eq-WeakComp} is easy to implement numerically, as will next be illustrated on the example of the stacked SSH model. Fig.~\ref{fig-StackedSSHWeak} shows the innocent looking density of states of $L^{w,j}_{\kappa,\rho}$ with many low-lying eigenvalues. The half-signature is for parameters as stated is $24$. Hence $\Ch_{\{1\}}(A)\approx \frac{24}{69}$. Furthermore one indeed finds numerically $\Ch_{\{2\}}(A)\approx 0$.  

\section{Bulk-boundary correspondence in semimetals}

The physical relevance of the weak winding numbers $\Ch_{\{j\}}(A)$ discussed above is rooted in the bulk-boundary correspondence by which they dictate the surface density of the flat band of edge states of the half-space Hamiltonian $\widehat{H}$ with Dirichlet boundary conditions. More precisely,
\begin{equation}
\label{eq-BBC}
\widehat{\Tt}(\sigma_3 \Ker(\widehat{H}))
\;=\;
\cos(\alpha)\,\Ch_{\{1\}}(A) 
\:+\;
\sin(\alpha)\,\Ch_{\{2\}}(A) 
\;.
\end{equation}
Here $\alpha$ is the (possibly irrational) cutting angle of the half-space, $\widehat{\Tt}$ is the trace per surface along the boundary and finally $\sigma_3$ is the Pauli matrix in the grading of the chiral Hamiltonian as in \eqref{eq-OdLocDef}. The robust equality \eqref{eq-BBC} holds under the above conditions assuring the existence of $\Ch_{\{j\}}(A)$ \cite{SSt} and thus for all parameters of the Hamiltonian. Let us note that it is an extension  to semimetals and to arbitrary $\alpha$ of similar identities for topological insulators \cite{PSbook}. For the case of clean graphene and a rational angle $\alpha$, the identity \eqref{eq-BBC} has been found before \cite{DUM}. \Red{The corresponding edge states in graphene have been observed experimentally \cite{TaoEtAl,AtaEtAl}.} Let us also note that the kernel of $\widehat{H}$ typically lies in one chiral sector so that the left-hand side of \eqref{eq-BBC} is simply $\pm \widehat{\Tt}(\Ker(\widehat{H}))$ with a sign pending on which chiral sector is occupied.  Finally let us stress that \eqref{eq-BBC} also holds for chiral Hamiltonians in arbitrary dimension \cite{SSt}. Furthermore similar identities linking weak Chern numbers to surface current densities in $d=3$ Weyl semimetals can be derived \cite{MT}.

\begin{figure}[t]
\includegraphics[width=1.0\linewidth]{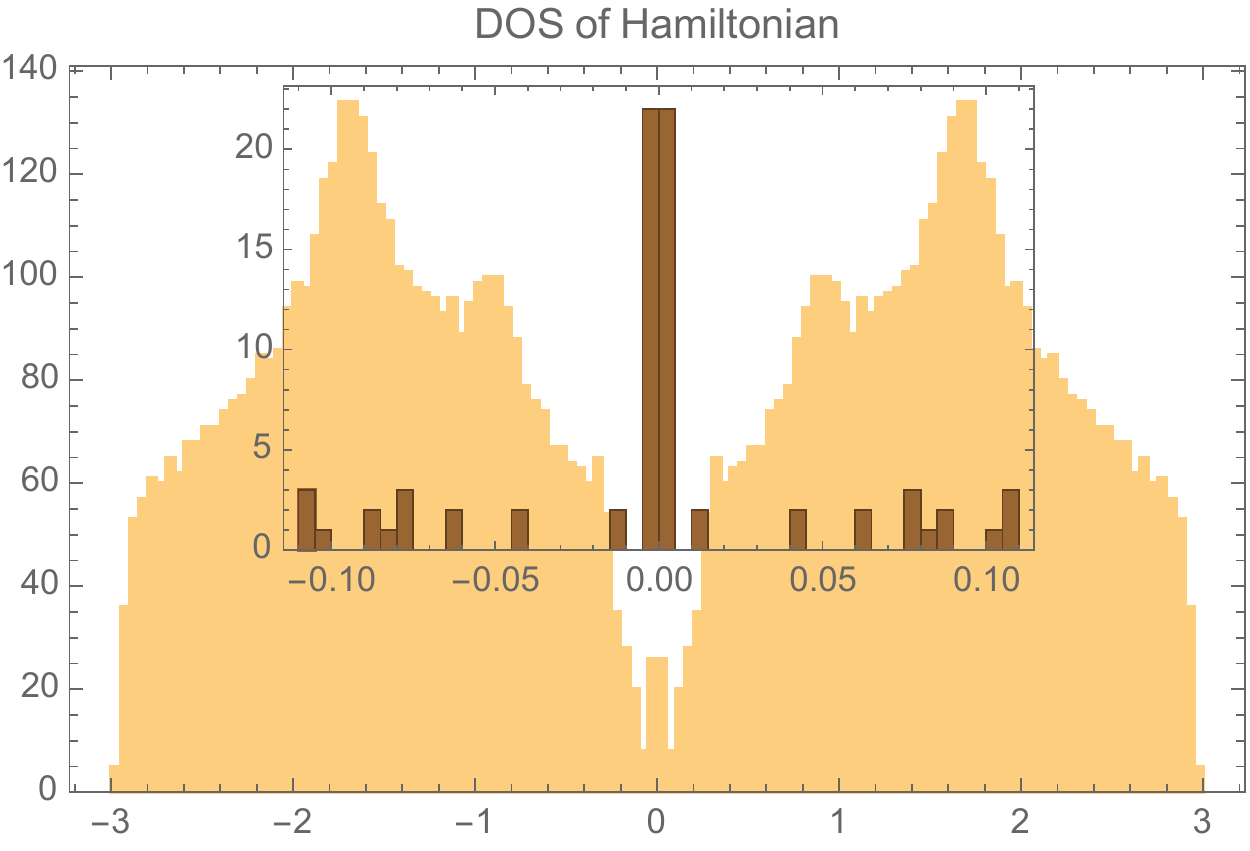}
\caption{Histogram of the eigenvalues of one realization of a stacked random SSH Hamiltonian restricted to $[-\rho,\rho]^2$ with Dirichlet boundary conditions. All parameters are as in Fig.~\ref{fig-StackedSSH}. The central flat band contains $48$ states.}
\label{fig-StackedSSHFlat}
\end{figure}

Based on \eqref{eq-BBC} and the hypothesis that all zero modes lie in one chiral sector, this allows to predict the total number of surface states lying in the flat band for a finite volume Hamiltonian $H_\rho$ on $[-\rho,\rho]^2$ with Dirichlet boundary condition. The square has $4$ edges, two of which (in the $2$-direction) have no surface states and two of which (in the $1$-direction) have about $(2\rho+1)\Ch_{\{1\}}(A)\approx 24$ surface states for the parameters exactly as in Fig.~\ref{fig-StackedSSHWeak}. Thus the total number of zero energy surface states should be equal to $48$. This is confirmed by counting the states in the flat band in Fig.~\ref{fig-StackedSSHFlat}.

\section{Summary and outlook}
The spectral localizer allows to make sense of Dirac and Weyl points in disordered media and is an efficient tool for their numerical computation, as well as for the numerical evaluation of other (weak) invariant in disordered semimetals. Let us also list several interesting open questions going beyond the results of the present paper. Is it possible to use a suitable spectral localizer to compute the \Red{number of Dirac or Weyl points} of the surface spectrum of a topological insulator? If one views \eqref{eq-DWcount} as way to count the number of points on the Fermi surface, is it possible to use the spectral localizer also to compute the density of states of a larger Fermi surface? What are the spectral statistics of the spectral localizer and how are they connected to properties of the physical system?

\acknowledgments
Both authors acknowledge financial support by DFG SCHU 1358/6-2, T.~S. also of the Studienstiftung des Deutschen Volkes.

\bibliographystyle{unsrt}

\end{document}